\documentclass[
aps,
pra,
longbibliography,
superscriptaddress,
 amsmath,amssymb,
twocolumn,
]{revtex4-1}

\usepackage{graphicx}
\usepackage{dcolumn}
\usepackage{bm}
\usepackage{upgreek}
\usepackage{color}
\usepackage{hyperref}
\DeclareMathAlphabet{\mathpzc}{OT1}{pzc}{m}{it}
\usepackage{enumitem}

\newcommand{\m}{\mathrm}

\newcommand{\eref}[1]{Eq.~(\ref{#1})}
\newcommand{\fref}[1]{Fig.~\ref{#1}}

\usepackage[normalem]{ulem}

\setlist[enumerate]{itemsep=-1mm}

\begin{document}

\title{Ground-state cooling of a mechanical oscillator by heating}



\author{Cheng Wang}
\affiliation{Department of Applied Physics, Aalto University, FI-00076 Aalto, Finland}

\author{Louise Banniard}
\affiliation{Department of Applied Physics, Aalto University, FI-00076 Aalto, Finland}

\author{Kjetil Børkje}
\affiliation{Department of Science and Industry Systems, University of South-Eastern Norway, PO Box 235, Kongsberg, Norway}

\author{Francesco Massel}
\affiliation{Department of Science and Industry Systems, University of South-Eastern Norway, PO Box 235, Kongsberg, Norway}

\author{Laure Mercier de L\'epinay}
\affiliation{Department of Applied Physics, Aalto University, FI-00076 Aalto, Finland}

\author{Mika A. Sillanp\"a\"a}
 \email{Mika.Sillanpaa@aalto.fi}
\affiliation{Department of Applied Physics, Aalto University, FI-00076 Aalto, Finland}
%

%

\date{\today}


\begin{abstract}




Dissipation and the accompanying fluctuations are often seen as detrimental for quantum systems, since they are associated with fast relaxation and loss of phase coherence. However, it has been proposed that a pure state can be prepared if external noise induces suitable downwards transitions, while exciting transitions are blocked. We demonstrate such a refrigeration mechanism in a cavity optomechanical system, where we prepare a mechanical oscillator in its ground state by injecting strong electromagnetic noise at frequencies around the red mechanical sideband of the cavity. The optimum cooling is reached with a noise bandwidth smaller than, but on the order of the cavity decay rate. At higher  bandwidths, cooling is less efficient. In the opposite regime where the noise bandwidth becomes comparable to the mechanical damping rate, damping follows the noise amplitude adiabatically, and the cooling is also suppressed.

\end{abstract}

\maketitle


\emph{Introduction.}---Recently, many breakthroughs have been accomplished in preparing and measuring macroscopic and non-classical motional quantum states. These achievements enable many potential applications, such as ultra-sensitive measurements and quantum information processing taking advantage of acoustic modes. These mechanical systems also provide an ideal platform to investigate a pristine quantum system interacting with its environment. 

Sideband cooling via interaction with an electromagnetic degree of freedom is a powerful technique to cool quantum systems, starting from small systems such as trapped ions \cite{Wineland1989CoolGnd,Chuang2009SBcool} or nanoparticles \cite{Aspelmeyer2013levitate,Barker2015levitate,Novotny2019levitate,Aspelmeyer2020Levit}, all the way up to nearly macroscopic mechanical resonators \cite{Aspelmeyer2006cool,Heidmann2006,Marquardt2007,WilsonRae2007,Kippenberg2008cool,Schwab2010,Teufel2011b,AspelmeyerCool11,Steele3D,Nakamura3D,Polzik2021entangle,2022_SiNBAE,Schliesser2022SiN}. In the latter, ground-state cooling can be obtained relatively easily in a cavity optomechanics setup. Its implementation relies on coherently pumping the lower-frequency motional sideband of a cavity mode, promoting the coupling between motion and electromagnetic fields through the radiation pressure force. The interaction can be seen as reservoir engineering, where the  oscillator is coupled to an effective cold bath, which mitigates and eventually dominates at large coupling rates the oscillator's fluctuations coming from its hot intrinsic thermal bath. Related ideas are  used to stabilize quantum squeezing \cite{SchwabSqueeze,Squeeze,TeufelSqueeze,Lehnert2019Squeeze} and entanglement \cite{4BAE,Teufel2021entangle} of mechanical oscillators. Engineered dissipation can then be used as a resource: state collapse upon measurements has been studied for usage in quantum computation \cite{Verstraete2009}.

Here, we answer an even more intriguing question on dissipation and noise: can fluctuations, such as those arising from a hot bath, be used to prepare a mechanical system in a pure quantum state? Mari \emph{et al.} \cite{Mari2012CoolHeat} have theoretically proposed to cool a mechanical oscillator in an optomechanical cavity by introducing the coupling to another, very hot, cavity mode. In their scheme, the hot mode establishes an interaction similar to that in sideband cooling between the oscillator and the cold cavity. Related studies have been presented \cite{Cleuren2012,Mari2015HeatHeat,Du2016NVcenterHeat}, including a connection to thermal machines \cite{Meystre2014engine,Wu2021HeatEngine}. In the present work, we realize the proposal of Ref.~\cite{Mari2012CoolHeat}, but in a manner recently discussed in Ref.~\cite{Naseem2021coolheat}. Instead of using another cavity mode, we introduce noise at suitable frequency components of the same mode. Thereby, we investigate the relevance of the spectral content of the energy used in sideband-cooling experiments, conjecturing that cooling is obtained whether the driving power comes in the form of a coherent tone, or in the form of broadband noise.


\emph{Theory.}--- 
Let us consider a cavity optomechanical system.
The mechanical oscillator of frequency $\omega_m$ and  intrinsic damping rate $\gamma$ is coupled to an environment at a temperature $T$, corresponding to the equilibrium thermal occupancy $n_m^T = \left[ \exp (\hbar \omega_m /k_B T) -1 \right]^{-1}$. The oscillator's position is denoted by $x(t) = x_{zp} (b^\dag + b)$, with $x_{zp}$ the zero-point motion amplitude, and with the phonon creation and annihilation operators $b^\dag, b$. The electromagnetic cavity has a frequency $\omega_c$ and a damping rate $\kappa$, and is described by the photon operators $a^\dag, a$. The oscillator is coupled parametrically to the cavity through the interaction $H_c = \hbar g_0 a^\dag a (b^\dag + b) $. Here, $g_0$ is the single-photon radiation-pressure coupling in angular frequency units.

\begin{figure*}[t]
   \centering
   \includegraphics[width=16cm]{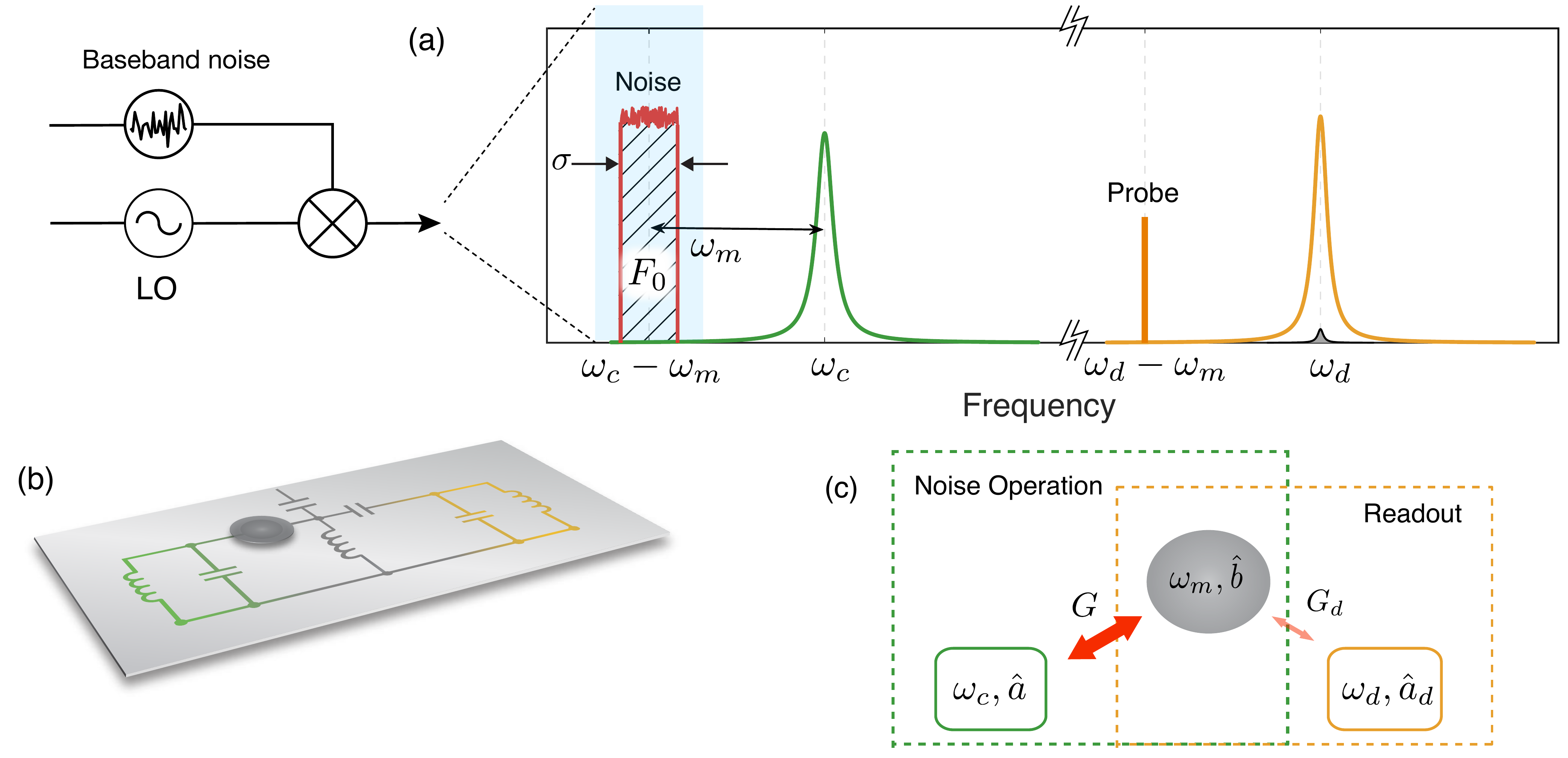} 
    \caption{\textit{Device and noise injection setup.} (a) Band-limited microwave noise is created by upconverting MHz-frequency band-pass filtered noise up to the red mechanical sideband center frequency. (b,c) Electromechanical device, which supports a drumhead oscillator and two microwave cavity modes, which are used for the noise driving (frequency $\omega_{c}$), and for independent monitoring (frequency $\omega_{d}$) of the oscillator. Two cavity modes are defined by the combination of Al microstrip meander inductors and the moving drum capacitors.} 
    \label{fig:figure1}
\end{figure*}

Let us first review the familiar sideband cooling with strong coherent drives, where the interaction can be linearized. In these protocols, the cavity is pumped with a single tone at a frequency close to the lower, ``red'' sideband, inducing a photon number $n_0$ in the cavity. The linearized interaction becomes $H_c = \hbar G (a^\dag +a) (b^\dag + b) $, where the effective coupling is $G = g_0 \sqrt{n_0}$. It is possible to show that the oscillator experiences an added optical damping equal to $\Gamma_\mathrm{opt} = 4 G^2/\kappa$ in the resolved-sideband limit $\omega_m \gg \kappa$. 
As this additional damping is almost not accompanied by enhanced fluctuations, the oscillator is effectively cooled down to a thermal occupancy
\begin{equation}
\label{eq:nm}
n_m = \frac{\gamma \, n_m^T +\gamma_\m{opt} \, n_{ba}}{\gamma_\m{eff}}  \,,
\end{equation}
where the effective linewidth of the oscillator is $\gamma_\m{eff} = \gamma + \gamma_\m{opt}$, $\gamma_\m{opt}$ is the added damping, and $n_{ba}$ is a small correction due to a non-ideal sideband-resolution \cite{Marquardt2007,WilsonRae2007} --negligible in our experiment-- or a classical occupation of the cavity mode. If sideband-cooling is the only process modifying the effective linewidth in the experiment, then $\gamma_\mathrm{opt} = \Gamma_\mathrm{opt}$.


The cooling due to radiation-pressure force can also be understood according to the description known as the quantum noise approach \cite{Marquardt2007,GirvinReview,Ojanen2014,OptoReview2014,BowenBook}. Let the oscillator be exposed to a force $F(t)$, which will introduce an energy term $H_F(t) = - F(t) x_{zp} (b^\dag + b)$. The damping rate of the oscillator contributed by the force becomes $\gamma_\m{opt} = \frac{x_{zp}^2}{\hbar^2} \left[ S_{FF}(\omega_m) -S_{FF}(-\omega_m) \right]$, where $S_{FF}(\omega)$ is the spectral density of the force. Thus, the damping is given by the difference of the spectral densities at positive and negative frequencies. This is reminiscent of the absorption-emission asymmetry in quantum mechanics, however, the frequencies are now expressed in a rotating frame, and such an imbalance can also be constructed classically.

In the case of cavity optomechanics, the force is the radiation-pressure force $F(t) =- \frac{\hbar g_0 }{x_{zp}} a^\dag a$. The damping is written  $\gamma_\m{opt} = g_0^2 \left[ S_{\delta n \delta n}(\omega_m) - S_{\delta n \delta n}(-\omega_m)\right]$ and can be interpreted as being due to fluctuations of the photon number $n$ ($S_{\delta n \delta n}(\omega_m)$ is the corresponding spectral density). The frequencies are here given with respect to $\omega_c$. Thus, if we can construct a photon number fluctuation spectrum very asymmetric with respect to the cavity frequency, significant cooling reminiscent of the sideband cooling is expected.

We now implement ``cooling by heating'' by introducing a fluctuating drive  into the cavity. With a view to describing our experiment, the spectrum of this drive is modeled as a box function of width $\sigma$ centered at the red sideband as shown in \fref{fig:figure1}(a). The spectrum is created by injecting band-filtered white electromagnetic noise into the cavity, corresponding to a  flux of $F_0$ photons per second, which gives rise to the time-averaged total photon number $\bar{n}_0 \simeq \frac{\kappa F_0}{\omega_m^2}$ in the cavity. For details, see the Supplemental Material \cite{supplement}. The damping becomes
\begin{equation}
\label{eq:goptnoise}
\gamma_{\m{opt}} = \frac{4 g_0^2 F_0}{\omega_m^2} \frac{\kappa}{\sigma} \tan^{-1} \left( \frac{\sigma}{\kappa}\right) \,,
\end{equation}
and the phonon number is given by \eref{eq:nm}. For a given total photon flux, if the noise bandwidth is larger than $\kappa$, cooling will be modestly suppressed. However, if the noise bandwidth is small, $\sigma/\kappa \ll 1$, this analysis indicates that we recover the sideband-cooling result $\gamma_{\m{opt}} \Rightarrow \Gamma_{\m{opt}}$ with $G = g_0 \sqrt{\bar{n}_0}$, which would stay valid down to zero noise bandwidth. This seems to agree with the intuition that a given electromagnetic input flux applied in the same frequency range could power the same cooling. However, the issue is more complicated as will be discussed next. 

\begin{figure*}[t]
   \centering
   \includegraphics[width=18cm]{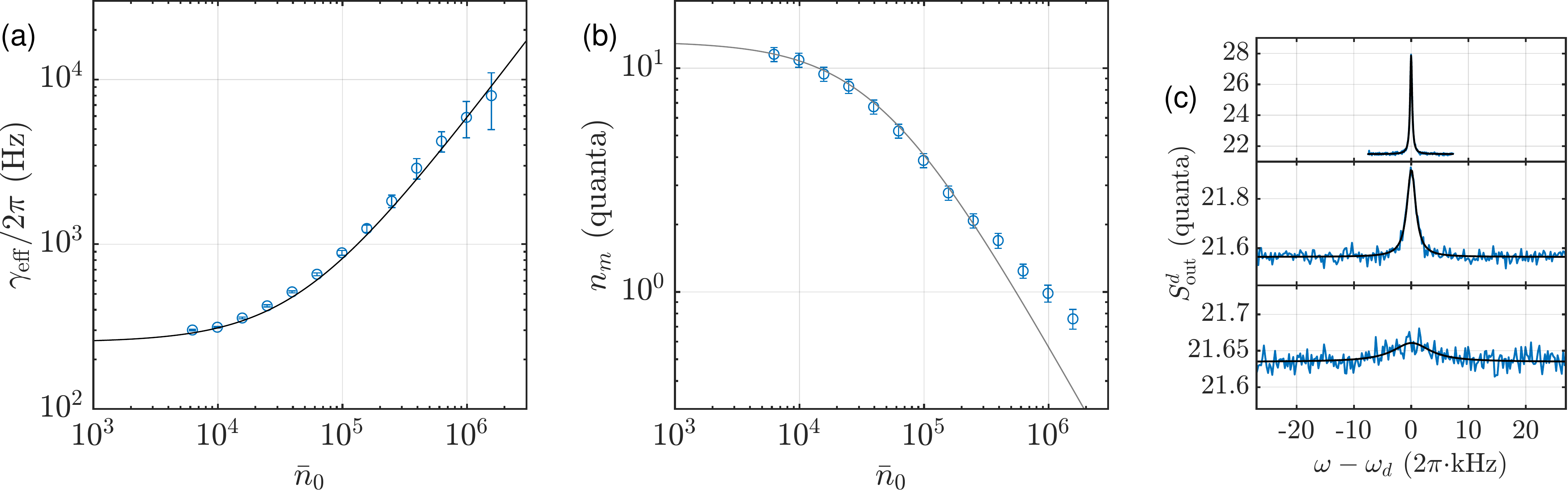} 
    \caption{\textit{Ground-state cooling with noise.} The noise bandwidth is kept constant at $\sigma/2\pi = 200$ kHz, while the incoming photon flux is varied. (a) The oscillator's effective linewidth, and the theoretical prediction based on \eref{eq:goptnoise} expressed as a solid line. (b) Phonon number of the oscillator. The solid line presents \eref{eq:nm}, with $n_m^T \simeq 27$ slightly enhanced by probe's technical heating. (c) Example probe spectra used for mode thermometry, at the photon numbers $\bar{n}_0 = [24.1 \times 10^3, 2.43 \times 10^5, 1.53 \times 10^6]$ from top to bottom.} 
    \label{fig:powersweep}
\end{figure*}

Let us now examine in more detail what happens if the noise bandwidth is of the order of $\gamma_\m{eff}$ or smaller. Now, the quantum-noise approach looses its validity, essentially because ensemble and time averages start to differ. In the intermediate regime $\gamma_{\m{opt}}/\sigma \approx 1$,
we can find perturbative corrections to the results from the quantum noise approach \cite{supplement}.
In the extreme limit $\gamma_{\m{opt}}/\sigma \gg 1$, the oscillator's damping adiabatically follows the slowly varying noise flux, giving a quasi steady state at all times.
We obtain that the cooling is, surprisingly, suppressed by a significant numerical factor in the latter limit. Physically, the reason is relatively simple: the occupation number is strongly dominated by instances of time when the cooling is weak, even though at other times the cooling is strong. 
Also, the profile of the time-averaged spectrum is very strongly dominated by periods of modest cooling associated to a narrow linewidth. In the limit $\gamma_{\m{opt}}/\sigma \gg 1$, $\gamma_{\m{opt}}/\gamma \gg 1$ the average phonon number and effective mechanical linewidth, as determined from the time-averaged spectrum, become
\begin{equation}
\begin{split}
\label{eq:lowsigma}
& n_m = \frac{\gamma n_m^T}{\gamma_\mathrm{opt}} \left[ \ln \left(\frac{\gamma_\mathrm{opt}}{\gamma} \right) - C  \right] \,, \\
& \gamma_\mathrm{eff} = \gamma  \ln \left(\frac{\gamma_\mathrm{opt}}{\gamma} \right) \,. 
\end{split}
\end{equation}
Here, $C$ is Euler's constant.



\emph{Experimental setup.} --- Our sample consists of a superconducting aluminum microcircuit which supports two microwave resonance modes (cavities), each coupled to two mechanical oscillators. The mechanical oscillators are drumhead membranes \cite{Teufel2011a,Teufel2011b} suspended above electrodes, forming displacement-dependent capacitors. For this experiment, we consider only one of the drums, and ignore the other owing to a significant frequency separation.  A schematic of the circuit design is shown in \fref{fig:figure1} (b).


The oscillator of interest is a 17 $\mu$m diameter aluminum drumhead resonating at ${\omega_m/2\pi \simeq 9.22\,\rm MHz}$ with an intrinsic damping rate $\gamma/2\pi \simeq 120$ Hz. The cavity mode with the frequency $\omega_{c}/2\pi \simeq 4.87$ GHz is used as a ``pump'' cavity to implement cooling, as it has the larger electromechanical coupling ${g_{0}/2\pi \simeq 39\,\rm Hz}$, allowing for stronger cooling effects. The higher-frequency, ``probe'' cavity with the corresponding parameters ${\omega_{d}/2\pi \simeq 6.42\,\rm GHz}$ and $g_{0,d}/2\pi \simeq 34$ Hz, is used for characterizing the oscillator thanks to calibrated thermometry. The coupling of the oscillator to the two cavities is schematically displayed in \fref{fig:figure1} (c). Both cavity modes satisfy the condition of resolved sideband limit where the respective damping rates of the pump and probe cavity $\kappa/2\pi \simeq 1.06$ MHz and $\kappa_{d}/2\pi \simeq 0.84$ MHz are much smaller than the mechanical frequency.




The device is studied at a temperature of $\sim 10$ mK in a dilution refrigerator. We find that, in the absence of active cooling, the mechanical mode is thermalized down to $T \simeq 10.6$ mK, where we denote the equilibrium thermal population as $n_m^T = n_{m,0}^T \simeq 24$.

Before describing the actual cooling-by-heating experiment, let us present the thermometry principle. We drive the probe cavity with a weak coherent tone at the red sideband, inducing the effective coupling $G_d$ (\fref{fig:figure1} (c)). This drops a thermal sideband at $\omega_{d}$ reproducing the mechanical spectrum. However, one has to bear in mind that the probe causes backaction itself, in this case residual sideband cooling, renormalizing $\gamma$ and $n_m^T$. We include all this in the thermometry calibration carried out by regular sideband cooling analysis via the probe cavity, which allows to associate a mechanical occupancy to a spectral density of electromagnetic flux in the thermal sideband. We select a probe power that is as small as possible but still allows for a detectable signal within hours of integration. This corresponds to a probe cooperativity $\mathcal{C}_d = \frac{4 G_d^2}{\kappa_d \gamma} \simeq 2.2$ and a cooling by a factor of two of the oscillator by the probing. Nonetheless, the overall cooling will be fully dominated by the noise-induced cooling.





The noise with the tunable bandwidth $\sigma$ is generated first in the MHz frequency range \cite{supplement}. Digitally-generated and -filtered pseudo-random numeric values are played at a rate of 100 MHz by an FPGA module to produce a voltage noise waveform with a box-function spectral profile centered at 7 MHz.  This noise is then up-converted by mixing it with a local oscillator (LO) at ${\omega_{c}-\omega_m - 7\,\rm MHz}$. We thus obtain a sharply band-pass-filtered noise whose box-shaped spectrum is centered at the red sideband, with the original bandwidth $\sigma$. 
After additional analog filtering and processing (see SM for a complete description of the processing \cite{supplement}), the noise injected is a good approximation to bandpass filtered thermal noise.

Based on the room-temperature characterization of the injected noise, we cannot directly obtain a prediction for $\gamma_{\m{opt}}$, since $F_0$ in \eref{eq:goptnoise} is difficult to determine experimentally at the sample, owing to the instrumental attenuation and noise separating the sample and the room-temperature equipment. We calibrate the product $g_0^2 F_0$ using the sideband cooling effect of a standard coherent tone applied to the pump cavity, where we also utilize the ratio of total noise power versus the coherent tone power which we directly measure at room temperature \cite{supplement}.




 

\begin{figure}[t]
   \centering
   \includegraphics[width=0.9\linewidth]{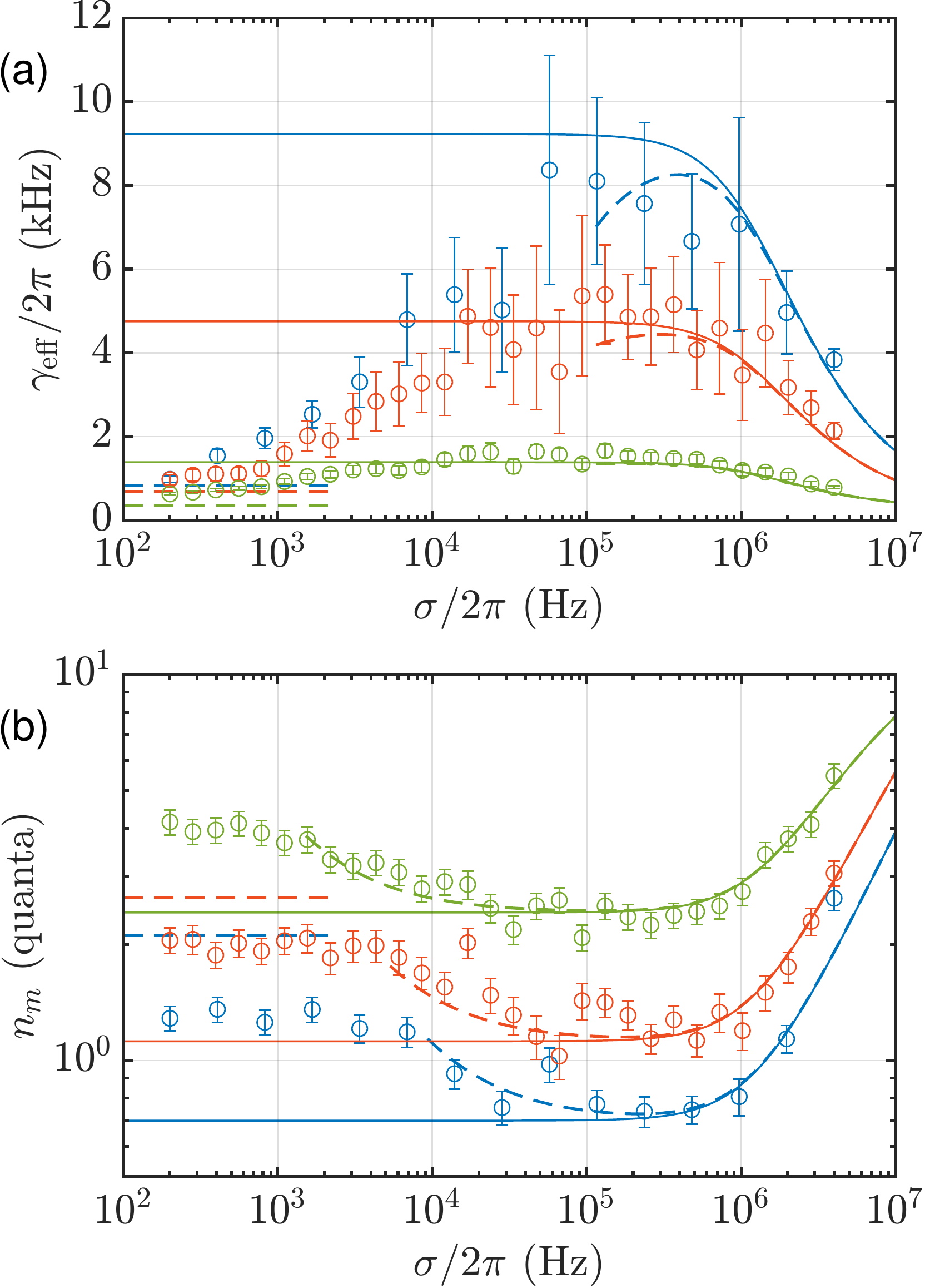} 
    \caption{\textit{Noise bandwidth dependence of cooling.} (a) Linewidth of the probe spectrum profile, and (b) Phonon occupancy, as functions of noise bandwidth. The photon numbers are $\bar{n}_0 = 1.57 \times 10^6$ (blue), $\bar{n}_0 = 7.9 \times 10^5$ (red), and $\bar{n}_0 = 2.0 \times 10^5$ (green). The solid lines are based on \eref{eq:goptnoise} and \eref{eq:nm}. The dashed curves depict corrections to quantum noise approach when $\sigma$ approaches $\gamma_\m{opt}$ from above. Horizontal dashed lines depict low-bandwidth predictions,  which are not shown for the lowest flux as \eref{eq:lowsigma} is not valid in this regime.  The theoretical curves are color coded according to the experimental data.} 
    \label{fig:BWsweep}
\end{figure}

\emph{Experimental results.} --- We now select a noise bandwidth $\sigma/2\pi = 200$ kHz, which is expected to yield a cooling close to optimum, and vary the integrated noise power. The effective damping shown in \fref{fig:powersweep} (a) follows closely the theoretical prediction which does not involve adjustable parameters. In \fref{fig:powersweep} (b), we show ``ground-state cooling by heating'', where we obtain mechanical occupancy $n_m \simeq 0.77 \pm 0.08$, integrated from the probe spectra as shown in \fref{fig:powersweep} (c). The theoretical prediction in (b) describes a situation where the intrinsic baths of the mechanics and of the cavity retain their natural occupancies, that is, where the system does not suffer from ``technical heating'' induced by the applied microwave power. As seen, there are deviations from this model towards the largest noise flux, which is mostly attributed to heating of the mechanical bath, here up to $n_m^T \simeq 60$ quanta. Such heating, here quite modest, is typically observed in microwave-optomechanical experiments, and remains unexplained thus far. Thanks to careful analog processing of the noise, we are confident that there is no significant noise leakage, i.e., heating by $S_{\delta n \delta n}(-\omega_m)$ remains negligible.

Finally, in \fref{fig:BWsweep} we investigate how the cooling performance depends on the noise bandwidth, while the total noise flux is kept fixed. We vary the bandwidth over 4 decades, spanning from $\sigma \approx \gamma \ll \gamma_\m{opt}$ up to $\sigma > \kappa \gg \gamma_\m{opt}$. This way, we can examine the expected suppression of cooling in both extreme limits. The effective linewidth of the probe signal is shown first in \fref{fig:BWsweep} (a) at different noise flux values. As discussed above, the low-bandwidth value is not of a strong physical significance, but reflects that the spectrum is dominated by times with little cooling. The occupancy shown in \fref{fig:BWsweep} (b) instead represents the time-averaged energy of the mechanical oscillator. The theory lines in \fref{fig:BWsweep} employ only calibrated parameters, and no free parameters. The suppression of cooling in the large-bandwidth limit is well captured by the theory, however, there is a numerical discrepancy between the measured occupancy and the result of the theoretical model employed at low bandwidths. We believe this can be due to time-correlations in the repetitive noise signal introduced by our noise-generation method in a higher proportion for small-bandwidth noise.


 


\emph{Conclusions.}--- In summary, we have cooled a mechanical oscillator close to its ground state by a noisy electromagnetic field.
Our method could be extended to prepare not only the ground state, but more sophisticated quantum states in mechanical systems relying on heat and fluctuations. If one would assign a temperature to the electromagnetic noise flux, the highest noise flux employed in the present experiment would be on the order of $10^8$ Kelvins. Such a level of Johnson noise can clearly not be created by simply heating a resistor. However, in a different system with a much  larger coupling $g_0$ \cite{LSETNEMSexp}, this temperature would be in the Kelvin range, and, intriguingly, one could use suitably filtered ambient noise for quantum state preparation.

%
%

\begin{acknowledgments} We would like to thank Matthijs de Jong for useful discussions. We acknowledge the facilities and technical support of Otaniemi research infrastructure for Micro and Nanotechnologies (OtaNano). This work was supported by the Academy of Finland (contracts 307757, 312057), by the European Research Council (101019712), and by the Finnish Cultural Foundation. The work was performed as part of the Academy of Finland Centre of Excellence program (project 336810). We acknowledge funding from the European Union's Horizon 2020 research and innovation program under grant agreement 824109, the European Microkelvin Platform (EMP), and QuantERA II Programme (13352189). L. Mercier de Lépinay acknowledges funding from the Strategic Research Council at the Academy of Finland (Grant No. 338565). FM acknowledges financial support from the Research Council of Norway (Grant No. 333937) through participation in the QuantERA ERA-NET Cofund in Quantum Technologies.
\end{acknowledgments}


%

\end{document}